\documentclass[runningheads]{llncs}
\usepackage{graphicx}
\usepackage[table,xcdraw]{xcolor}
\usepackage{multirow}
\usepackage{subfig}

\usepackage{moreverb,url}
\usepackage{svg}

\usepackage[colorlinks,bookmarksopen,bookmarksnumbered,citecolor=red,urlcolor=red]{hyperref}
\usepackage{amsmath}
\usepackage{natbib}
\newcommand\BibTeX{{\rmfamily B\kern-.05em \textsc{i\kern-.025em b}\kern-.08em
T\kern-.1667em\lower.7ex\hbox{E}\kern-.125emX}}

\usepackage{url}

\usepackage{breakurl}

\usepackage{moreverb,url}
\usepackage[colorlinks,bookmarksopen,bookmarksnumbered,citecolor=red,urlcolor=red]{hyperref}

\usepackage{hyperref}
\usepackage{algorithm}
\usepackage{algorithmic}
\usepackage[normalem]{ulem}
\usepackage{natbib}
\useunder{\uline}{\ul}{}

\usepackage{graphicx}
\usepackage[table,xcdraw]{xcolor}
\usepackage{multirow}
\usepackage{subfig}

\usepackage{url}

\begin{document}


\title{Trends of Pink Slime Journalism Advertisement Expenditure and Spread on Facebook from 2019-2024}

\author{Christine Sowa Lepird \inst{1} \and
Lynnette Hui Xian Ng \inst{1} \and
Kathleen M. Carley \inst{1}
}

\institute{
Carnegie Mellon University, USA \\
\email{csowa@andrew.cmu.edu}
}




\maketitle  
\begin{abstract}
Pink slime journalism is a practice where news outlets publish low-quality or inflammatory partisan articles, claiming to be local news networks. This paper examines the spread of pink slime sites on Facebook using public posts from Pages and Groups. We evaluate the trends of sharing pink slime sites on Facebook and patterns regarding the advertisements purchased by the parent organizations of the pink slime news networks. Our analysis discovers that while the number of pink slime posts on Facebook pages have decreased over the years, advertising dollars have increased. The increase in advertising dollars influences an increase in Facebook group posts. Further, the advertising expenditure increases during election years, but contentious topics are still discussed during non-election years. By illustrating the patterns and themes from US election years of 2020,  2022, and 2024, this research offers insights into potentially dangerous journalism tactics, and provides predictions for future US Presidential Elections.
\end{abstract}


\section{Introduction}

Pink Slime journalism is a practice that refers where organizations masquerade as local news outlets, and publish low-quality, hyperpartisan or inflammatory articles \citep{moore_consumption_2023}. These organizations typically have no ties to the local community and are owned by a parent organization that operates a network of news outlets to push ``local" news targeting multiple states. This practice is common in America, where the term ``local" represents a region in the United States that is a size of a state or smaller. Pink slime news sites are characterized by algorithmically-generated websites, with visually similar content, layouts and origins \citep{burton2020research}. The same message is pushed to multiple states, while its tone is tweaked to appeal to the local community.

To exploit the trust in local reporting, parent organizations like Metric Media have created almost 1,000 local news sites \citep{bengani_metric_2021}. Local reporting remains a highly trusted source of news, although there is a dearth of authentic news reporting. The creators of pink slime news networks take advantage of this trust and publish low-quality news on their websites, often on politically related and partisanship issues. Besides solely publishing stories on news websites, these pink slime journalism networks have created social media accounts in which they share the URL (Unified Resource Locator) link of the pink slime news on social media platforms. In this article, we study the activity and patterns of sharing of pink slime websites on Facebook. Facebook was chosen because it is a social media platform that is commonly used for sharing URLs to pink slime news sites \citep{moore_consumption_2023}.

The activity of publishing partisan news provides the possibility for manipulation of public opinion, and more so if these pink slime news networks are highly funded by advertising money. This leads us to our first research question: \textbf{RQ1: How much money was spent by pink slime organizations to target which populations?} Through this research question, we look at advertising dollars spent in total on Facebook advertisements (ads) by the larger parent organizations, and the breakdown of the target advertisement demographics. Through textual analysis methods, we examine the posts on the Facebook groups that shared pink slime news URLs. This investigation answers our second research question: \textbf{RQ2: How did the ad spend influence online conversations?}

News sites from metric media are active in publishing partisan news, providing a possibility of manipulation of public opinion. Although there is a dearth of authentic local reporting by local reports, it remains highly trusted. The creators of these networks are taking advantage of this trust, and publish low-quality news on these sites, often politically related partisanship. Pink Slime news organizations are typically active throughout the year, but since they primarily share news with a political slant, we examine one last research question: \textbf{RQ3: How have pink slime tactics differed over time of a political event, in particular the United States (US) elections?} 

With this study of pink slime news URLs on Facebook, we examine targeting patterns of low-quality local news and the advertising money poured into the pink slime journalism ecosystem. Our studies provide insights towards the evolving roles of journalism in the digital age, where news websites and social media platforms overlap in their dissemination and broadcasting functions.

\subsection{Contributions}
Although there are studies point to the use of generated news sites for political issues \citep{burton2020research}, overall, the success of pink slime journalism on social media platforms is understudied. In this study, we make use of content and geographical analysis to understand how the production strategies of pink slime organizations have evolved on Facebook. We study the production and distribution of pink slime URLs across three key political events in the United States: (1) the 2020 Presidential Elections, (2) 2024 Presidential Elections, and (3) the 2022 Midterm Elections. 

In this study, we make the following contributions:
\begin{enumerate}
    \item We analyze the advertisement expenditure of pink slime news organizations per state across political election timelines, allowing for projection of advertising dollars poured into political campaigns.
    \item To facilitate our analysis, we collected a dataset from Facebook containing posts that have pink slime URLs and advertising expenditure from the 2020 to 2024 United States (US) Elections. This dataset provides a representation of pink slime journalism on Facebook in the context of the US Elections, thereby facilitating future studies of this phenomenon.
    \item Finally, we analyze the messaging techniques of pink slime news organizations per state across political election timelines.
\end{enumerate}

\section{Background}
\subsection{Local News Journalism}
Local news agencies have faced many challenges in the 21st century. The newspaper industry, which is predominantly local newspapers, has experienced a 67\% decline in advertising revenue since its peak in 2005 \citep{pewresearchLocalNews}. Between 2008 and 2018, newsrooms experienced a 25\% reduction in employment, with newspapers seeing a much higher 47\% loss of labor \citep{brookings}. Since 2004, over 2,000 local newspapers have closed, resulting in 200 US counties having no local newspaper.

Americans have higher trust in local news than in national news organizations \citep{gottfried_partisan_2021}. 71\% of US adults (falsely) believe their local news outlets are doing well financially despite only 14\% of these adults having donated to these outlets in the past year \citep{pewresearchLocalNews}. To cope with the falling revenue, some regional newspapers have been acquired by larger corporations, forming corporate-owned local news sites, which thus have to publish content to be shared across multiple locales in a region \citep{LeBrun2022,Toff2021}. These acquisitions result in decreased local content publications and increase national news coverage. 

Some news outlets, like the sports-focused \textit{The Athletic}, thrived by implementing a national framework to report on local events \citep{Ferrucci2022}. Further, these news outlets suffer from a dearth of local reporters to write stories, so newsrooms have turned to Artificial Intelligence (AI) to assist in writing news stories in various fashions. While some of the examples of AI usage are designed to automate tasks (like the \textit{Los Angeles Times'} QuakeBot which automatically drafts an article if the U.S. Geological Survey detects an earthquake), other usages can be more sinister and fully generate ``local" stories with no local context \citep{Quere2022_Trust}.

\subsection{Pink Slime Journalism}
In this age of digital media, digital journalism production evolved to include pink slime journalism, which was initially described in 2005 as automated news reporting \citep{cohen2015pink}. Given the particular financial and labor challenges the local newsroom has faced since then, five organizations have adopted this genre of news with a goal of creating ``local" news using algorithms and a handful of non-local reporters. Media companies face declining profits and expanding demands for content, thus pink slime journalism rely on digital tools to generate content and publish more for less, reducing the quality of the news \citep{moore_consumption_2023}.

The largest of the most prominent organizations, Metric Media, controls vast swaths of pink slime sites that do not appear to have foreign ties \citep{bengani_metric_2021}, but are currently financed by political candidates and political action committees with the hope of swaying election results. When speaking of threats to election integrity, Alex Stamos, director of the Stanford Internet Observatory, remarked ``The issue [...] is not going to be foreign interference. It's much more likely that legitimate domestic actors possibly operating under their own name — with LLCs or corporations with very shady funding that are not required to disclose what that funding is — are going to dominate the online conversation about the outcome of the election" \citep{murphy_local_2020}. Metric Media has been shown to algorithmically generate most of its content, and it prioritizes the publication of state and national partisan content at the expense of local news \citep{royal2022local}. One such example of algorithmically generated content is to use an Application Programming Interface (API) to display the local weather. The API is a piece of computer script that retrieves the local weather from a weather source via only a zip code. The local weather is a news topic which American adults rate as the most important local news topic for their daily lives \citep{pewresearchLocalNews}, and can be easily generated via computer programming.

Further, research from Stanford University analyzed news consumption of pink slime sites and found that at least 3.7\% of American adults visited at least one pink slime site during the 2020 Presidential elections period \cite{moore_consumption_2023}. The frequency which pink slime sites are visited reflects the extent which these sites can have an effect on population opinion, especially during crucial events like the elections. Social media posts by these pink slime websites have been observed to implement information maneuvers against local and state elections. In contrast, authentic local news sites focused on the national elections and state figureheads \citep{lepird2023comparison}.

\subsection{Spread of News Through Facebook}
Around 70\% of Americans get their news from the social media platform Facebook \citep{gramlich_10_nodate}. 15\% of the adults cited social media as their preferred platform for reading local news, while 13\% of them prefer print newspapers \citep{pewresearchLocalNews}. Those who prefer reading their local news via social media are less likely to be closely following the news stories than those who consume local news via television or print \citep{pewresearchLocalNews}. Social media platforms like Facebook have been a serious threat to the traditional local news publishing industry. Local news organizations previously held exclusive access to their readers, however, in today's journalism climate, Facebook knows the intended audience of news websites through their micro-targeting tactics \citep{brookings}. When local news organizations post to Facebook, they have a greater incentive to optimize for engagement metrics. Therefore, these organizations post content pertaining to \textit{national} news stories as opposed to local stories \citep{Toff2021}. 

Despite the scale of news advertisements on Facebook, not all of this news is coming from quality news sources: 15\% of referrals to fake news sites are coming from Facebook \citep{guess_exposure_2020}, and 17.7\% of visits to pink slime sites are referred by Facebook \cite{moore_consumption_2023}. Many local news organizations have official pages on Facebook to share their stories with their social media followers. During the COVID-19 pandemic, researchers analyzed the posts of over 1,000 local news sites to Facebook Pages and found that the engagement on these posts rose proportionately with the size of the population the local news organization was targeting \cite{LeQur2022}.

In general, US citizens rate posts from local Facebook groups to be more trust worthy than non-local Facebook groups. Posts from Facebook pages of local news are perceived to be more interesting and more trustworthy to local citizens as compared to posts from local neighborhood Facebook groups \citep{Quere2022}. Much like authentic local news outlets, many of the known sources of pink slime have associated social media accounts on platforms like Facebook to amplify the spread of the messaging to the community. The names of these pink slime sites frequently contain the target community in the domain name. Examples are: The \textit{Bucks County} Standard, \textit{North Alaska} News, the \textit{Michigan} Star, and \textit{Pensacola} Times.  Per \citet{mihailidis_2021}, there is a serious reason to be concerned about neighborhood Facebook groups: the platform makes it difficult to discern who owns the group. Even if the group advertises to be apolitical, they may still have a political slant. This results in further political manipulation, as \citet{mihailidis_2021} expresses: ``A fractured local media ecosystem, especially without the investigative power once shared by healthy local newspaper competition, has left local-and-state politics open to further manipulation by outsiders. Local and state governments draw considerable attention because they hold more power than the average citizen acknowledges".

Local news organizations boost their presence on social media not only by frequent postings but also by advertisements. Meta allows for companies to purchase advertisements on Facebook. Newsrooms can have their stories promoted to their intended audiences on Facebook, while micro-targeting the desired age, locations and gender of the readership. During the 2018 US Midterms elections, \citet{Haenschen2022} ran ads in Texas to examine if the advertisements affect voter turnout. They found that only one set of targeted ads focusing on abortion rights and women's healthcare in a competitive district was shown to significantly increase turnout.

In 2018, Meta launched the Facebook Ad Library for increased transparency on the organizations that are running ads on Facebook. This library provides information in terms of how much money the organizations are spending and what messages they are promoting \footnote{\url{https://www.facebook.com/ads/library}}. This resource has allowed researchers to analyze the spread of advertising of posts with false information and political posts, observing that micro-targeted ads reduced the likelihood of persuading Democrat respondents to vote for Democrat candidates \citep{Liberini2023,Mejova2020}. 

\section{Data Collection}
Facebook is the social media platform with the plurality of referrals to pink slime sites \citep{moore_consumption_2023}, hence this paper studies the spread of pink slime sites on Facebook. We study how the sites share pink slime news through both organic and paid means. A total of three datasets were collected with respect to pink slime domains: (1) public posts on Facebook groups and Facebook pages (termed ``Facebook posts dataset"), and (2) total number of advertisement expenditure (ad-spend) per group (termed ``Advertisement expenditure").

\paragraph{Facebook posts dataset} We collected this data using the Python Crowdtangle API\footnote{\url{https://pypi.org/project/PyCrowdTangle/}}. For each pink slime domain listed in \citet{noauthor_towcenterpartisan-local-news_2023}, we used the API to collect all the posts that have a link from public Facebook groups and pages from 2019 until August 2024, which was when the CrowdTangle platform was shut down. Overall, this data collection yielded 1,108,059 posts from the pages and 43,357 posts from groups.

Facebook page posts in this dataset are dominated by posts made by official Facebook pages. These pages are dedicated to sharing news from individual pink slime sites. Therefore, data reflecting pink slime on Facebook pages can be seen as a glimpse into the marketing efforts by the organizations. The presence of Facebook groups that shared news from these websites indicates an organic news spread within local communities. These groups have a median subscriber count of 1,075 group members, and frequently contain local community names within their group names. In total, we collected posts from 12,644 Facebook pages, linking to 531 unique pink slime domains; and posts across 8,528 Facebook groups, linking to 198 unique pink slime domains.

From this Facebook posts dataset, we extract and quantified the number of pink slime domains per organization. \autoref{tab:orgs2} shows the breakdown of news domain present in this dataset. The largest organizations, Metric Media and Star, are known for their politically conservative leanings while Local Report, Courier, and American Independent have liberal political leanings.


\begin{table}[]
\centering
\begin{tabular}{|c|c|c|}
\hline
\textbf{Parent Organization}  & \textbf{Sub-Specialty} & \textbf{Websites} \\ \hline
\multirow{6}{*}{Metric Media} & Metric Media           & 398               \\ \cline{2-3} 
                              & Metro Business         & 51                \\ \cline{2-3} 
                              & LGIS                   & 34                \\ \cline{2-3} 
                              & Record                 & 11                \\ \cline{2-3} 
                              & Franklin Archer        & 10                \\ \cline{2-3} 
                              & Local Labs             & 8                 \\ \hline
Star                          &                        & 11                \\ \hline
Local Report                  &                        & 9                 \\ \hline
Courier                       &                        & 9                  \\ \hline
American Independent          &                        & 7                 \\ \hline
\end{tabular}
\caption{he number of pink slime sites shared on Facebook in this study by parent organization.}
\label{tab:orgs2}
\end{table}

It should be noted that Facebook page posts in this dataset are dominated by posts made by official Facebook pages dedicated to sharing news from individual pink slime sites. For that reason, the page data can be seen as a glimpse into (free) marketing efforts by the organizations. The Facebook groups that were not operated by organizations and shared news from these pink slime websites indicate an organic news spread within local communities. These groups are not seeded with the websites by the official community, and the sharing patterns indicate the natural propagation of pink slime sites. These groups had a median subscriber count of 1,075 group members and frequently had local community names in their group names like `Republicans of the Palm Beaches', `Progressive Democrats of North Carolina', `Michigan Republicans/Democrats Debate Group', `Middle Georgia Young Democrats', `Philadelphia 8th Ward GOP', and `Republican Liberty Caucus of Indiana.'

\paragraph{Advertisement expenditure dataset} This dataset was collected using the Facebook Ad Library\footnote{\url{https://www.facebook.com/ads/library}}. The dataset consists of posts that pink slime parent organizations paid to promote on Facebook pages, the amount of money known pink slime organizations spent on advertisements posts per state, and the impressions garnered. This was collected by searching for political ads from the parent organizations: Courier's (``Cardinal \& Pine", ``Courier Newsroom", ``Courier Newsroom, Inc.", ``Floricuas", ``Granite Post", ``Iowa Starting Line", ``The 'Gander Newsroom", ``The Copper Courier", ``The Keystone", ``The Nevadan", ``UpNorthNews", ``The Keystone Courier"), Metric Media's (``Metric Media LLC", ``Franklin Archer", ``Local Government Information Services", ``The Record"), and American Independent's (``American Independent Media"). No ads were found from the Star and Local Report parent organizations. We collected 14,581 advertisement posts shared from 310 Facebook pages, linking to 383 unique domains, the majority of which are direct pink slime domains. 




\section{RQ1: How much money was spent by pink slime organizations to target which populations?}
Our first research question examines the advertising expenditure (ad spend) of pink slime organizations and their targeted populations. We make use of data from the Advertising Expenditure dataset to perform this investigation. \autoref{tab:ads_over_time} shows the aggregated ad spend of all the organizations from 2018 through 2024 and the total number of impressions those ads received. In terms of ad spending across the years, the election years of 2020 and 2024 observed a spike in ad spend, and subsequently the election year of 2022 observed the next highest ad spending. The total impressions on advertisements were highest during the election years of 2020, 2022, and 2024, indicating the appeal of political ads these pink slime organizations during these years.

\begin{table*}[]
\centering
\begin{tabular}{|c|c|c|c|}
\hline
\textbf{Year} & \textbf{Number of Ads} & \textbf{Total Impressions} & \textbf{Total Ad Spend} \\ \hline
2018 & 374   & 1,328,813   & \$21,713    \\ \hline
2019 & 1,367 & 16,091,317  & \$200,067   \\ \hline
2020 & 2,774 & 105,847,118 & \$1,118,763 \\ \hline
2021 & 1,735 & 14,787,133  & \$158,883   \\ \hline
2022 & 3,696 & 43,776,653  & \$731,752   \\ \hline
2023 & 1,404 & 18,664,300  & \$544,398   \\ \hline
2024 & 3,231 & 328,211,953 & \$7,952,185 \\ \hline
\end{tabular}
\caption{Pink Slime Facebook Ads Over Time}
\label{tab:ads_over_time}
\end{table*}

\begin{figure*}
    \centering
    \includegraphics[scale=0.45]{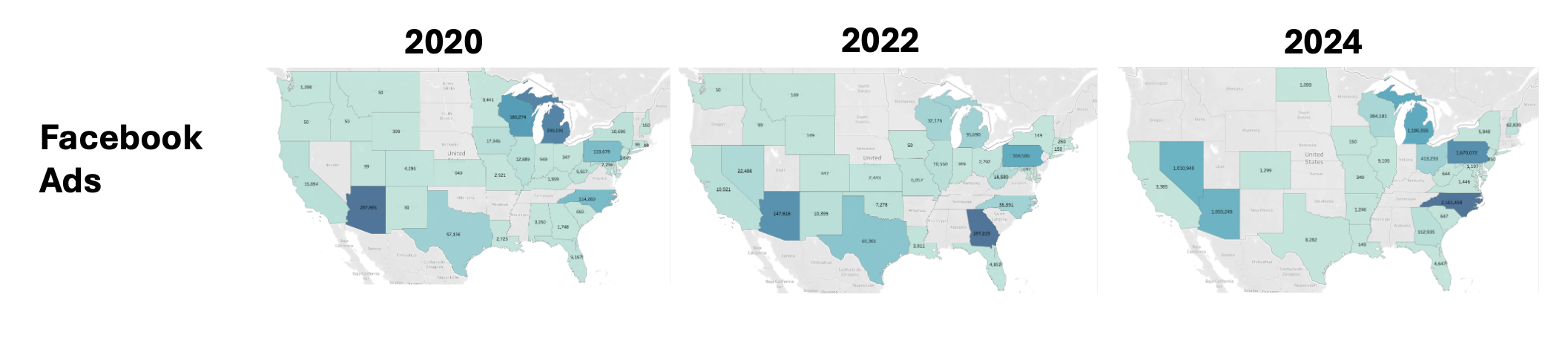}
    \caption{Total Pink Slime Advertising Expenditure by State During Election Years}
    \label{fig:ad_maps}
\end{figure*}

With the advertising expenditure segregated by states in \autoref{fig:ad_maps}, we analyze the expenditure per state. We observe that three of the five largest and known pink slime organizations purchased Facebook ads throughout the time period we examined, and Metric Media was the only organization to consistently purchase ads across all time periods. Courier purchased ads through 2020 then stopped doing so until December 2023, and American Independent purchased ads 2022-2024. All of the organizations that purchased ads targeted Pennsylvania - a key swing state that determined Joe Biden's 2020 victory over Donald Trump - as well as the swing states of Michigan, Wisconsin, and Georgia.

The top ten states by ad spend (from highest to lowest) are North Carolina, Pennsylvania, Arizona, Michigan, Nevada, Wisconsin, Ohio, Georgia, Texas, and West Virginia. All of these states with the exception of Ohio and West Virginia were among those with the closest 2020 Presidential Elections spread. Iowa had the 15th highest ad spend due to its unique position as the first state in the country to hold an electoral event with its caucus every four years. While swing states are an important target during election years, other states' significance was seen as a result of isolated events. For example, Courier spent \$16,999 in ads to Iowa in 2020. During the 2020 elections, Trump won Iowa against Biden by 8.2 percentage points. Given the small difference, it is likely that Courier, a left-leaning party, poured more resources to boost chances in Iowa.

When we assess the differences in the demographics of those targeted by the pink slime Facebook ads (\autoref{tab:ad-demographics}), we notice gender and age shifts between the organizations. Two politically left-leaning sites, American Independent and Courier Newsroom, target women more than men in their ads, and the majority of their impressions are coming from individuals under 55. However, Metric Media, a right-leaning organization, displays its ads to more men than women, and the majority  of their impressions come from Facebook users who are over the age of 55. It seems that left-leaning sites generally target younger women, while right-leaning sites generally target older men. 

\begin{figure*}
    \centering
    \includegraphics[scale=0.6]{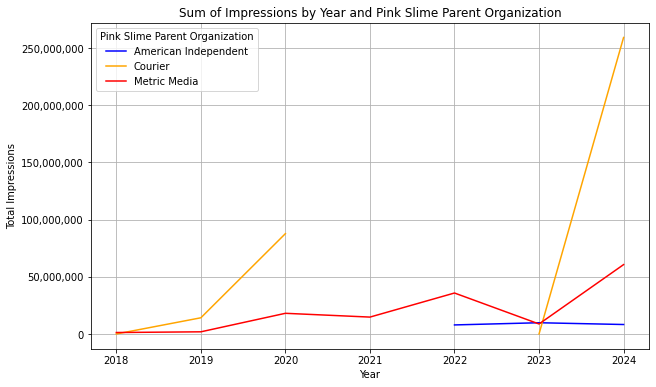}
    \caption{Advertising Expenditure by Parent Organization Over Time}
\label{fig:ad_by_parent_org}
\end{figure*}

\begin{figure}[!tbp]
  \centering
  \subfloat[An over-time plot of Page Posts Linking to Pink Slime Sites by Organization (through August 2024) \label{fig:pages_plot}]
  {\includegraphics[width=0.45\textwidth]{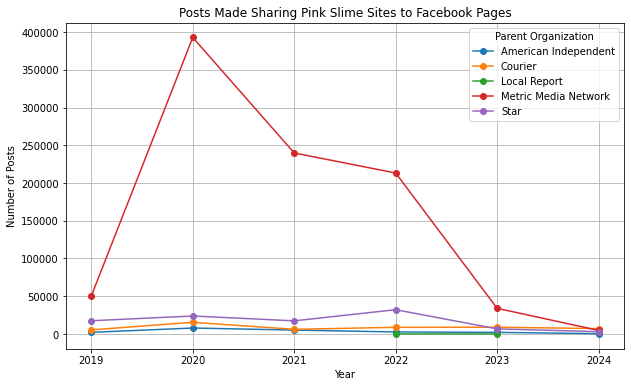}}
  \hfill
  \subfloat[An over-time plot of Group Posts Linking to Pink Slime Site by Organization (through August 2024) \label{fig:groups_plot}]
  {\includegraphics[width=0.45\textwidth]{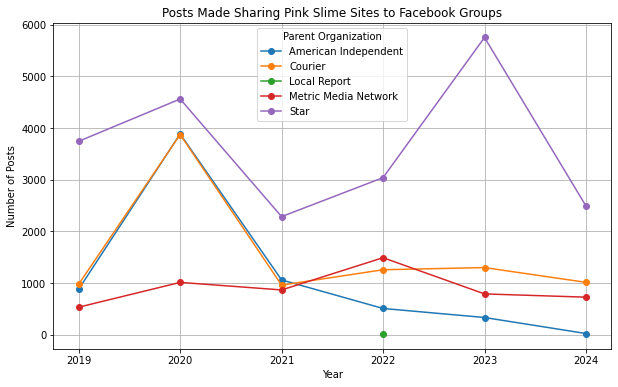}}
  \caption{Facebook Pages and Group Posts Over Time}
\end{figure}

\begin{figure*}
    \centering
    \includegraphics[scale=0.5]{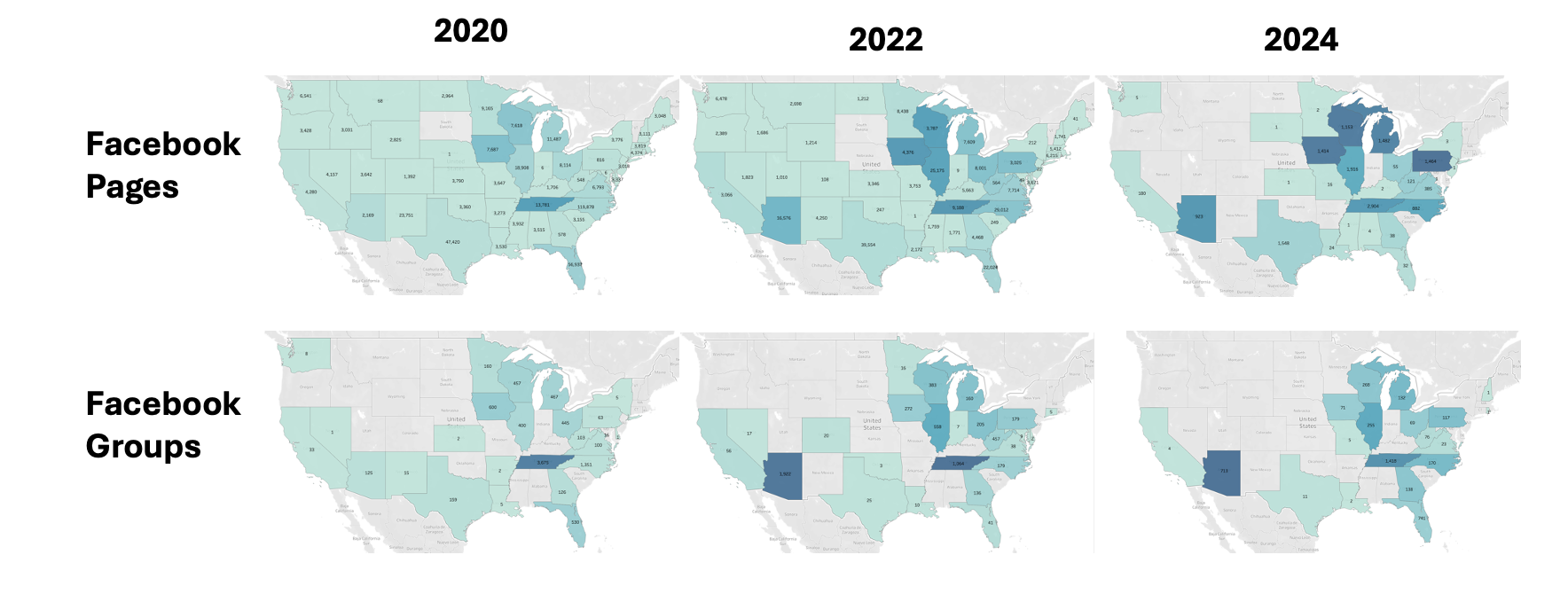}
    \caption{Facebook Pages and Groups Posts by State}
    \label{fig:groups_pages_elections_map}
\end{figure*}

\begin{table*}[]
\centering
\begin{tabular}{|l|l|l|l|l|l|l|l|l|}
\hline
                     & Female & Male   & 18-24  & 25-34  & 35-44  & 45-54  & 55-64  & 65+    \\ \hline
American Independent & 69.6\% & 29.8\% & 4.5\%  & 18.8\% & 22.6\% & 20.2\% & 17.7\% & 16.2\% \\ \hline
Courier Newsroom     & 57.2\% & 40.2\% & 14.3\% & 27.0\% & 20.3\% & 14.7\% & 11.4\%  & 10.5\%  \\ \hline
Metric Media         & 43.4\% & 49.8\% & 5.4\%  & 11.3\% & 10.3\% & 14.3\% & 22.5\% & 30.3\% \\ \hline
\end{tabular}
\caption{Breakdown of targeted ad demographics by gender and age for pink slime organizations}
\label{tab:ad-demographics}
\end{table*}

Finally, we examined whether state variables have an effect on ad spend. We performed Pearson Correlations of several state variables to the 2022 and 2024 pink slime ad spend. This includes the 2020 voter spread\footnote{https://www.presidency.ucsb.edu/statistics/elections/2020}, 2020 GDP\footnote{https://www.bea.gov/}, percentage of population living in rural areas\footnote{https://www.census.gov/en.html}, percentage of population with a bachelors degree\footnote{https://fred.stlouisfed.org/release/tables?rid=330\&eid=391444\&od=2020-01-01} and so forth. We used the statistics retrieved from the 2020 data, envisioning that organizations only have access to that information in 2022 and 2024. \autoref{tab:stats-correlations} presents the Pearson correlation coefficient between the state variables and the ad spend, and the corresponding p-values. We find that only the 2020 Voter Spread is significantly correlated with ad spend at the $p<0.05$ level.

\begin{table*}[]
\centering
\begin{tabular}{l|ll|ll|}
\cline{2-5}
                                                  & \multicolumn{2}{c|}{2022}           & \multicolumn{2}{c|}{2024}            \\ \hline
\multicolumn{1}{|c|}{Variable} &
  \multicolumn{1}{c|}{Correlation} &
  \multicolumn{1}{c|}{Significance} &
  \multicolumn{1}{l|}{Correlation} &
  Significance \\ \hline
\multicolumn{1}{|l|}{2020 Voter Spread}           & \multicolumn{1}{l|}{-0.52} & 0.0040 & \multicolumn{1}{l|}{-0.52}  & 0.0072 \\ \hline
\multicolumn{1}{|l|}{Cities Over 100k Population} & \multicolumn{1}{l|}{0.032} & 0.88   & \multicolumn{1}{l|}{-0.12}  & 0.57   \\ \hline
\multicolumn{1}{|l|}{\begin{tabular}[c]{@{}l@{}}Percent of Population\\ Living in Rural Areas\end{tabular}} &
  \multicolumn{1}{l|}{-0.085} &
  0.66 &
  \multicolumn{1}{l|}{-0.041} &
  0.84 \\ \hline
\multicolumn{1}{|l|}{\begin{tabular}[c]{@{}l@{}}Percent of Population \\ with a Bachelors Degree\end{tabular}} &
  \multicolumn{1}{l|}{-0.055} &
  0.78 &
  \multicolumn{1}{l|}{-0.12} &
  0.65 \\ \hline
\multicolumn{1}{|l|}{2020 GDP}                    & \multicolumn{1}{l|}{0.085} & 0.66   & \multicolumn{1}{l|}{-0.087} & 0.68   \\ \hline
\multicolumn{1}{|l|}{Median Age of State}         & \multicolumn{1}{l|}{-0.12} & 0.53   & \multicolumn{1}{l|}{0.12}   & 0.58   \\ \hline
\multicolumn{1}{|l|}{March 2020 Governor's Party} & \multicolumn{1}{l|}{-0.15} & 0.44   & \multicolumn{1}{l|}{0.34}   & 0.093  \\ \hline
\multicolumn{1}{|l|}{Electoral College Votes}     & \multicolumn{1}{l|}{0.15}  & 0.45   & \multicolumn{1}{l|}{-0.35}  & 0.87   \\ \hline
\end{tabular}
\caption{State variables and their Pearson correlation to 2022 and 2024 pink slime ad spend}
\label{tab:stats-correlations}
\end{table*}

\section{RQ2: How did the ad spend influence online conversations?}
In this research question, we combined data from advertising expenditure and posts content, and examined the effects of ad spend on online conversations. 

\begin{figure*}
    \centering
    \includegraphics[scale=0.52]{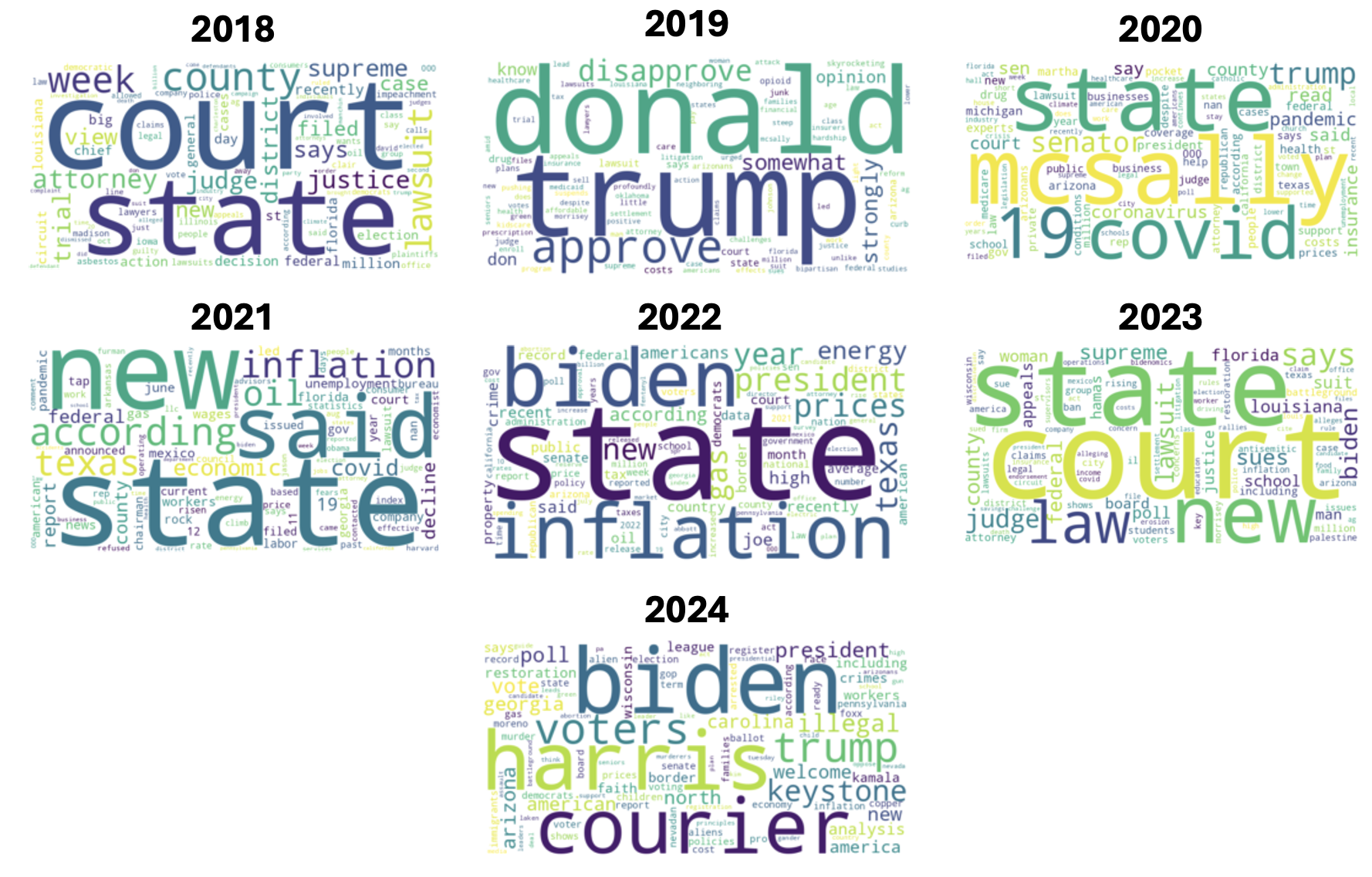}
    \caption{Wordclouds of the Top 100 Words Appearing in Pink Slime Facebook Ads Over Time}
    \label{fig:wordclouds}
\end{figure*}

\begin{figure*}
    \centering
    \includegraphics[scale=0.4]{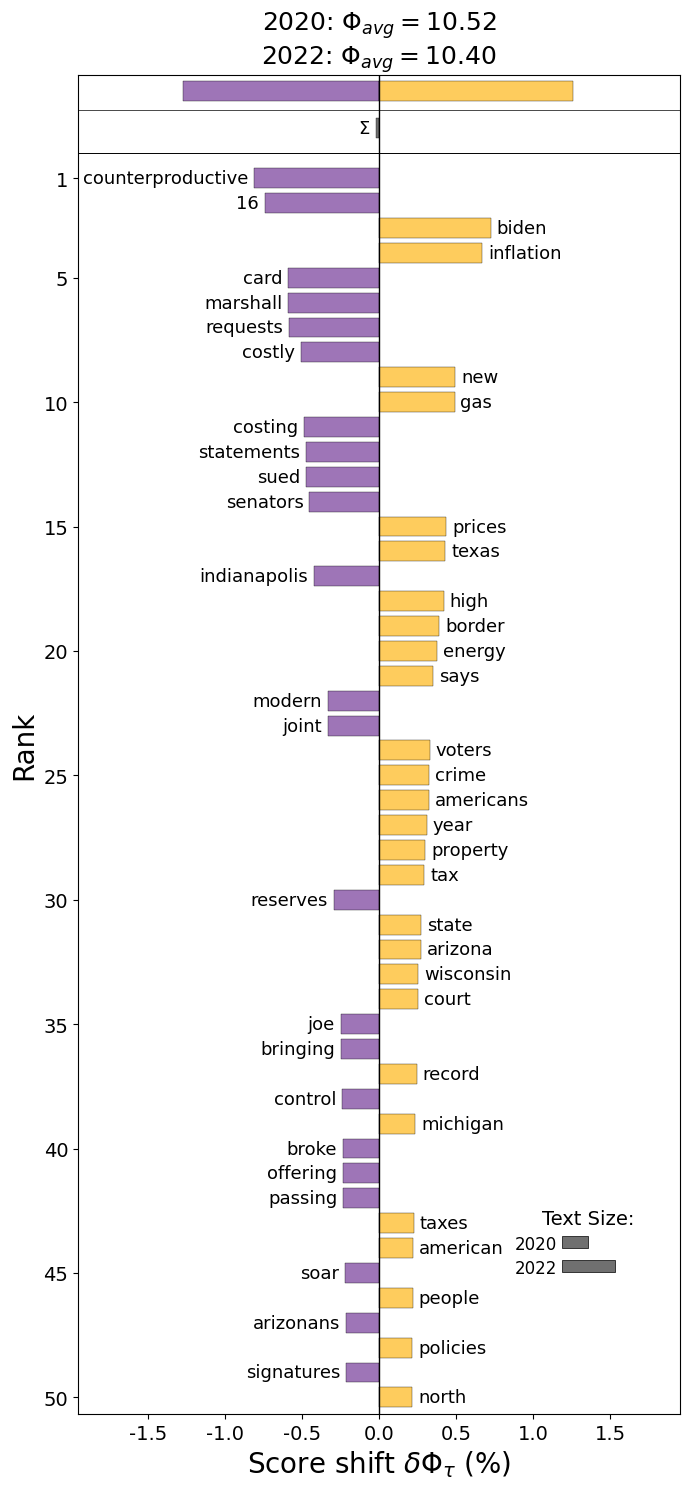}
    \caption{Change in words used in Facebook ads by Pink Slime Organizations in 2020 (left) and 2022 (right)}
    \label{fig:entropy}
\end{figure*}

We first begin by analyzing themes within the online conversations through a temporal lens. We broke down the content within Facebook ad messaging across the years and populated the frequency of messaging into a word cloud. To do so, we first pre-processed the text in the messaging to remove stopwords and URLs, before using Python's wordcloud package\footnote{\url{https://pypi.org/project/wordcloud/}} to formulate the word cloud of the top 100 words, sized by frequency of appearance. \autoref{fig:wordclouds} illustrates the changing focus of topics in election and off-election years and \autoref{fig:entropy} shows how these words changed between the two elections generated using \cite{Gallagher2021}. Throughout all of the years, the most targeted states (`Texas', `Michigan', `Georgia', `Arizona', `Florida', `Wisconsin', and more) remain as top terms. One key tactic used is to write ads with the same message but switching out the name of the state for the one that is being targeted in the ad. For example, some of the ads run in 2022 had the following titles:  ```Inflation has shot up a staggering 13.2\%' since Biden took office, Arizona's CPI at 13\%" , ```Inflation has shot up a staggering 13.2\%' since Biden took office, Michigan's CPI at 8.1\%", ``3 in 5 Americans concerned about housing affordability, North Carolina's average rent up 30\%", and ``3 in 5 Americans concerned about housing affordability, Wisconsin's average rent up 17\%."

During election years, ad spending increases drastically, and the conversations naturally turn political. While the presidential candidates' names were not at the forefront of the 2020 conversation, there was a focus on the phrase \textit{Catholic}. The ads containing references to `Catholic' were mostly run in September and October 2020 by the Metric Media organization, leading into the appointment of Catholic Supreme Court Justice Amy Coney Barrett. The top two of these ads, garnering 275,000 and 112,000 impressions, respectively, were titled `President Trump addresses Catholics directly' and `Catholic Vote: Biden's anti-school choice stance should worry WI Catholic school parents'; the `Catholic' phrase was indirectly used to support President Trump's re-election. 

During the midterms in 2022, President Biden was the top phrase, with secondary attention paid to key economic issues like `inflation', `gas' , and `prices'. Three of the top four ads (by impression) in 2022 were purchased by left-leaning American Independent, criticizing Georgia senate candidate, Herschel Walker, and garnering over 3 million impressions from 9 distinct ads. Right-leaning Metric Media, however, garnered over 350,000 total impressions with the ad title `As Pennsylvanians receive fourth stimulus check, Pigott points out negative real wage growth: 'Joe Biden is the pay cut president''. Much like the 2020 efforts to use the Catholic narrative into praise for President Trump, the 2022 tactic was to use negative economic news to undermine President Biden (and the Democratic party) ahead of the midterm election.

Ad expenditure during the years between elections drastically diminished, and the discourse focuses more on ``court". These ads highlight that state supreme and high courts, and are not necessarily political or partisan in nature. For example, `Appeals court vacates ruling against Parkways Authority over Turnpike toll fees' received over 22,000 impressions in 2023. This strategy may be to establish the Facebook Pages sharing the news as a nonpartisan, trustworthy local news outlet when they are not actively trying to push a political message, and to keep the organizations active and visible to the Facebook audience even during the down time.

\begin{figure*}
    \centering
    \includegraphics[scale=0.25]{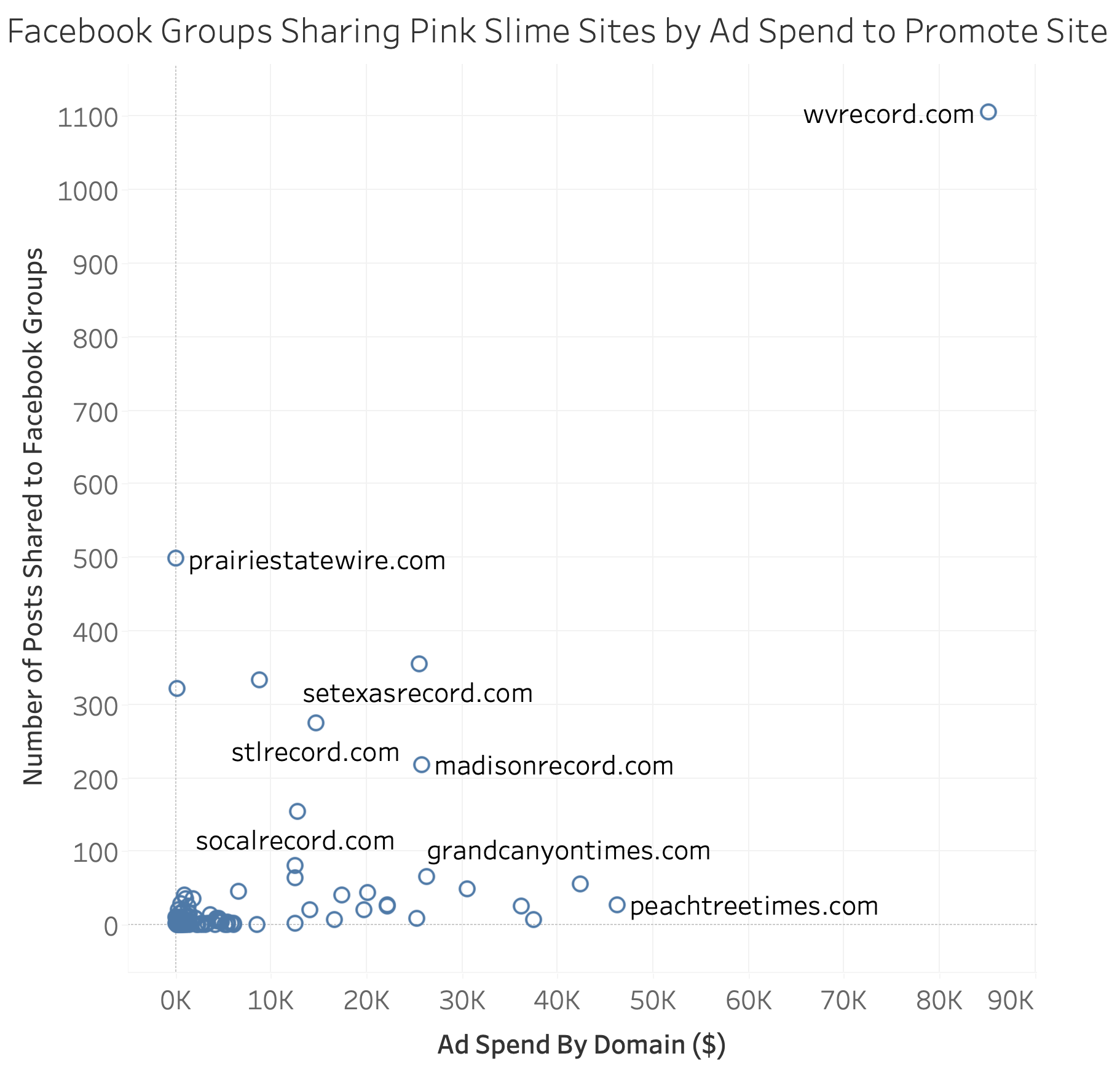}
    \caption{Number of Instances a Pink Slime Domain Appears in a Facebook Group by Ad Spend for those Domains}
    \label{fig:group_spend}
\end{figure*}

Finally, to see how the ad spend influenced organic conversation, we generate \autoref{fig:group_spend} to compare the amount of money spent promoting a given pink slime domain via Facebook ads and the number of times that domain appeared in a Facebook group. A linear trend line ($R^2 = 0.35$) indicates a weak positive correlation between the two measures, ad spend and organic conversation. 



\section{RQ3: How have pink slime tactics differed over time of a political event?}
Given data across multiple years, we finally ask how the different organizations have adapted their tactics during different political phases, namely two presidential elections, a midterm election, and between elections.

\autoref{fig:ad_by_parent_org}, \autoref{fig:groups_plot}, and \autoref{fig:pages_plot} show an over-time plot of the advertisement impressions, group posts linking to pink slime sites and page posts linking to pink slime sites per organization, respectively. In general, the advertising impressions increase drastically during election years, and decreases during the years in between elections. While advertisement impressions have increased over the years, the total number of posts in Facebook pages has decreased since 2020, indicating that pink slime organizations are now pivoting away from posting to their Facebook pages organically and instead spending more for paid advertisements to run from those pages. Facebook groups see different trends depending on the parent organizations. Star had peaks in 2020 and then again in 2023 while Courier and Local Report had peaks in 2020, and Metric Media peaked in 2022. All of these, with the exception of Star, show peaks in organic activity around election years (like ad impressions); however, since the 2024 data only contains data through August 2024, it's possible that Star would have exceeded its 2023 numbers had we had the ability to collect all of the 2024 calendar year data.

\autoref{fig:groups_map} shows the number of Facebook group posts that share pink slime websites affiliated with each state and \autoref{fig:pages_maps} shows the number of pink slime Facebook Page posts per state. Metric Media has the most prolific activity on Facebook posts, posting news stories relating to all states on pages. Metric Media owns hundreds of these pink slime sites whereas the other four organizations own fewer than 15 of these sites each. Their goal seems to be to reach \textit{every} corner of the United States with strong focus on Illinois, where the founder of Metric Media, Brian Timpone, resides. The other pink slime organizations have more concentrated sharing in Facebook groups and pages to states of higher electoral importance.

\paragraph{Metric Media}
Metric Media is the largest known parent organization of pink slime, and has several sub-networks associated with and sharing the IP space with main Metric Media sites \citep{bengani_metric_2021}.
Metric Media has spent more on ads (\$1,927,128.50) than American Independent but less than Courier. They started in 2019 and spent \$208,604.50 during the 2020 presidential election year. However, in 2022, they increased ad spend significantly, up to \$400,039 during the midterm year and final to \$977,253 during the 2024 presidential election year. Metric Media dominates the Facebook page post dataset with 392,849 posts in 2020. These numbers then rapidly decreased, with 213,249 in 2022 and just 4,735 in the first 8 months of 2024. This data coupled with the ad data indicates there was a business decision to pivot from aggressively posting to their Facebook pages towards spending more for ads. Facebook groups have posted their news sites 5,409 times with the highest year being in 2022 (1,489 posts) followed by 2020 (1,011 posts). This is consistent with past work which has also observed significant publications by Metric Media in Ohio, Michigan, Texas, California and Illinois \citep{royal2022local}.

\paragraph{Local Report}
Local Report is a new player in this space, having only posted on Facebook during the 2022 elections. The organization had no known ads purchased on Facebook. Unlike the other parent organizations, their 14 sites do not have a dedicated Facebook Page. Unsurprisingly, given their lack of social media accounts, they had minimal page posts - only 28, all in 2022 except for one in 2023. In groups, these sites have only been posted 12 times, all in 2022.

\paragraph{Star}
Like the Local Report, Star had no known ads purchased on Facebook, but unlike the Local Report, they have active Facebook pages. The 10 largest Facebook page sharers were the official social media accounts for these platforms and account for 85\% of the posts to Facebook pages that linked to their 11 news sites. The page posts increased from 2020 (23,826 posts) to 2022 (32,151 posts) but declined for the first 8 months of 2024 (3,035 posts). Star has the most group posts of any pink slime organization. Their news sites have been posted to groups 21,878 times; the highest posting frequency was in the year 2023 (5,755 posts).

\paragraph{Courier}
Courier  purchased Facebook ads in 2018 (\$495), 2019 (\$166,347) and 2020 (\$910,158.50) before going on a break and returning with vigor leading up to the 2024 presidential election by buying ads in December 2023 (\$996) and 2024 (\$6,909,696.50). They have seen a total of 51,933 posts from Facebook pages with the plurality coming in 2020 (15,460 posts). In 2022 they only had 8,829 page posts and for the first 8 months of 2024 they had 7,110 page posts. Courier has the second highest number of group shares - with a total of 9,376 posts to Facebook groups linking to their news sites; their biggest year for group shares was 2020 (3,875 posts).

\paragraph{American Independent}
The American Independent is the organization with the least ad spend, starting with purchasing ads in just 2022. In 2022 alone they spend \$331,713 on Facebook ads and they proceed to spend more on ads in 2023 (surprisingly, a non-election year) with \$415,990. Their 2024 ad spend fell to just \$65,235. There were 20,400 instances of Facebook pages and 6,682 instances of Facebook groups sharing links to the American Independent's news sites, the plurality of which occurred in 2020; however, the vast majority of these posts to Facebook groups and pages are to the American Independent parent site (not one of their smaller sites targeting swing states).

\section{Discussion}
In today's digital age, news reporting goes beyond traditional television and print newspapers. Journalists have begun to make use of social media to disseminate their news articles and broadcast their news, to keep up with the shifting readership trends. This gives rise to a new phenomenon: pink slime journalism. In this paper, we discuss the impact of pink slime journalism with respect to the US elections.

Understanding the impact of pink slime journalism through their propagation on Facebook is important for a few reasons. This understanding provides insight about the scope and the scale of pink slime networks throughout the US. While ad expenditure and group posts are focused on key states, page posts are dispersed across almost every state, indicating the prevalence of pink slime advertisements across America. 

In general, advertising expenditure from Pink Slime organizations have increased across the years. Metric Media is one of the largest players in the space. Metric Media spends heavily and broadly, producing ads that target a wide number of states and demographics. Other pink slime organizations produce ads that are more targeted towards swing states and vulnerable demographics. Courier was spending heavily on advertising to Iowa and to a lesser degree Michigan and Pennsylvania;  however, they seem to have discontinued their advertising expenditure after the 2020 Elections. In contrast, American Independent have started advertising in the 2022 Elections. 


In terms of messaging, during the election years, the focus of pink slime news websites were heavily on the politics during the elections. Between elections, these websites did not simply fizzle away, but kept up with talking about court cases and lawsuits. This is a technique to keep the Facebook accounts and pages active and fresh in people's memory.

Understanding the geographical trends of pink slime distribution in the 2020 to 2024 elections will be useful in providing an insight towards possible trends in future elections. From our studies, we believe that in future US elections, pink slime advertisement expenditure and Facebook posting will increase drastically. Metric Media will continue to spend heavily in an attempt to influence the election, and American Independent will continue with a focused and high ad spend on swing states. Finally, should another seat open up on the Supreme Court, the ads will target the religion and values of the president as a consideration.

\paragraph{Ethical considerations} Our data, while obtained from a social media platform, focused on the quantity of posts and ad-spend. Our study only uses public data and does not involve any personally identifiable information. We analyzed aggregate information, and not personal or individual information.

\paragraph{Future work} on pink slime journalism can link the production of pink slime as observed within our study with the consumption of pink slime, thereby further quantifying the impact of the pink slime news. This includes analyzing the pink slime phenomena to indicate the to reveal whether consumption habits matches production ideals, providing insights towards the success of journalism outlets. 

\section{Conclusion}
Pink slime journalism is increasingly being used as an advertising technique, in which politically leaning junk news appears to look like local news. In this work, we investigated the prevalence of pink slime journalism on Facebook groups and pages across five years from 2018 to 2024, characterizing their quantity across the states of America. Our observations show that pink slime news organizations have increased in their advertising expenditure across the years, and this has impacted organic conversation on social media. Pink slime journalism is a growing phenomenon in the US and the world, for they leverage on the public's trust of local news sites to dispense political news. Our work sets forth paths for analyzing this phenomenon across the US, in terms of messaging and advertising expenditure, opening avenues to examinations of this new journalistic trend.

\section{Acknowledgments}
This work was supported in part by the Office of Naval Research (ONR) Award N00014182106, the Knight Foundation, the Center for Computational Analysis of Social and Organizational Systems (CASOS), and the Center for Informed Democracy and Social-cybersecurity (IDeaS). The views and conclusions contained in this document are those of the authors and should not be interpreted as representing the official policies, either expressed or implied, of the ONR or the U.S. government.



\subsection{IRB Approval}

This research was conducted with IRB approval in the Spring of 2024 Federalwide Assurance No: FWA00004206 IRB Registration No: IRB00000603.

\bibliographystyle{plainnat}

\bibliography{aaai22_2}

\appendix

\section{Appendix A: Over Time Maps by Parent Organization}

\begin{figure*}
    \centering
    \includegraphics[scale=0.4, angle=90]{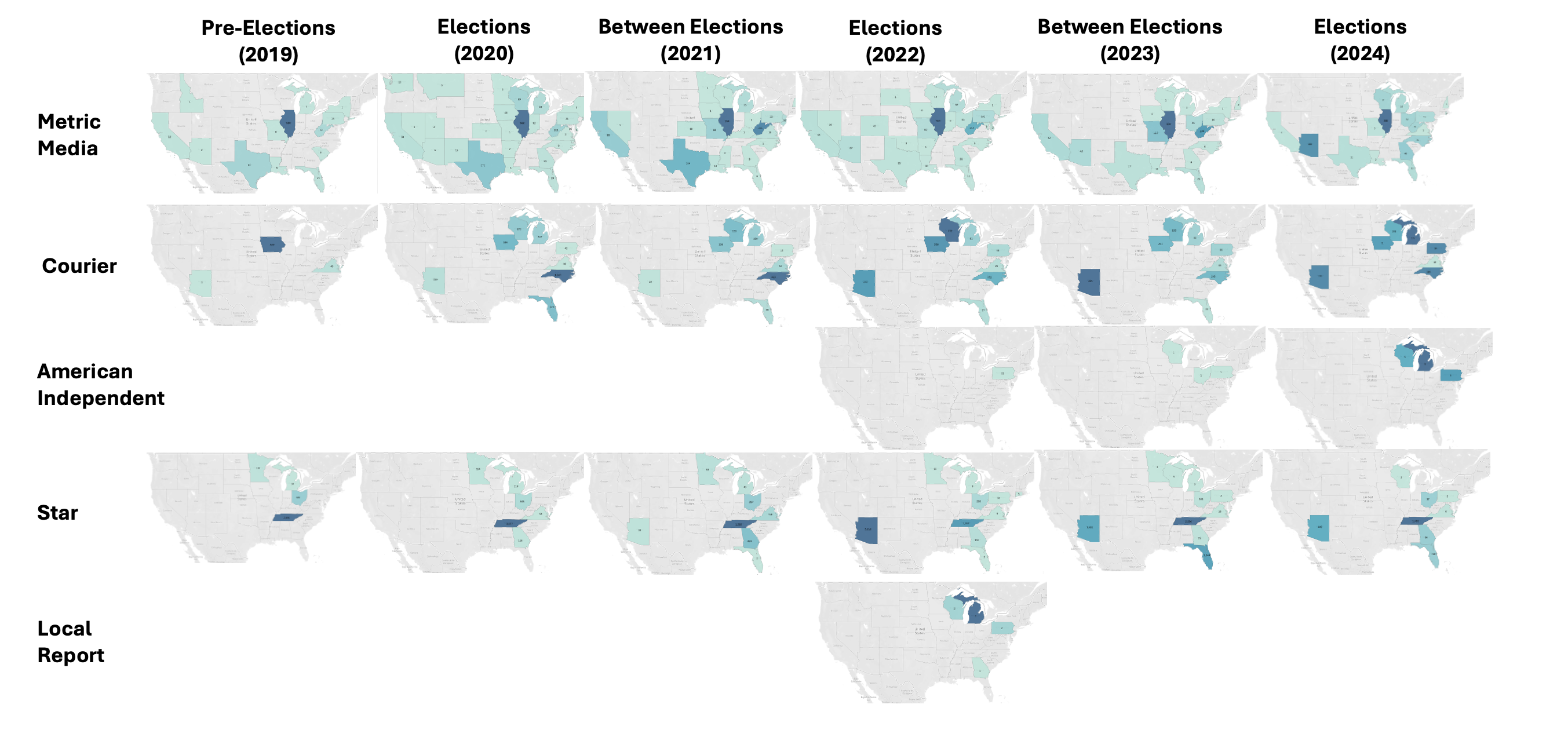}
    \caption{Sum of all the posts linking from public Facebook groups to pink slime sites targeting different states by year through August 2024.}
    \label{fig:groups_map}
\end{figure*}

\begin{figure*}
    \centering
    \includegraphics[scale=0.4, angle=90]{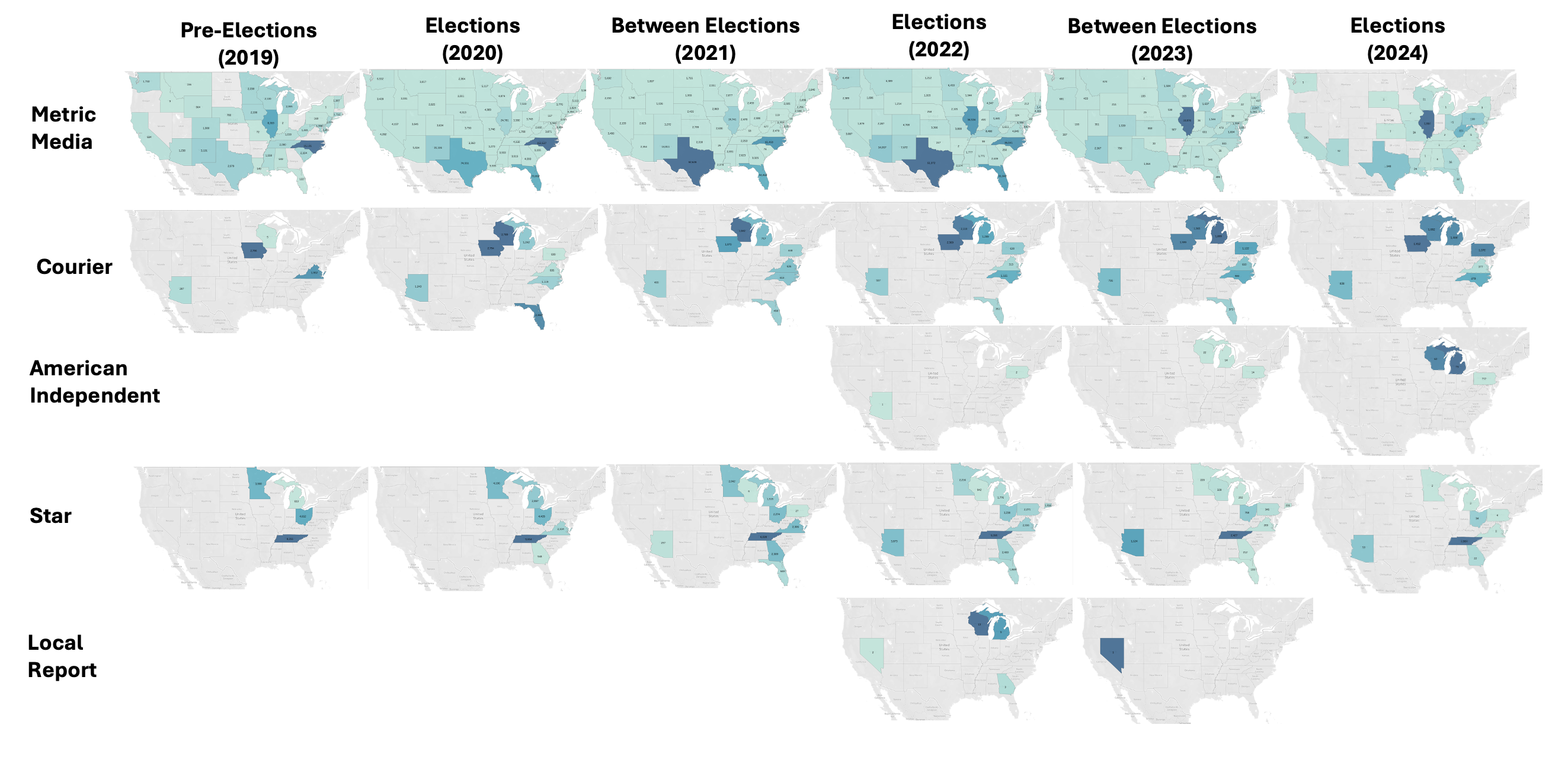}
    \caption{Sum of all the posts linking from Facebook Pages to pink slime sites targeting different states by year through August 2024.}
    \label{fig:pages_map}
\end{figure*}

\begin{figure*}
    \centering
    \includegraphics[scale=0.4, angle=90]{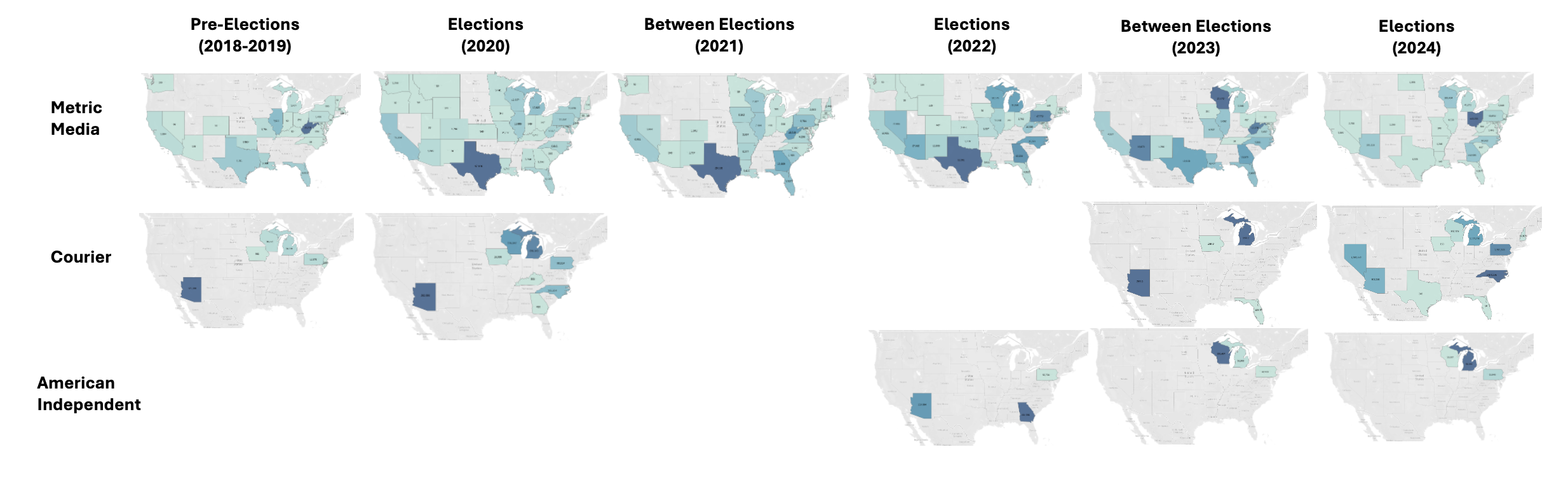}
    \caption{Total Facebook ad expenditure (in US Dollars) by state over time by the various pink slime organizations.}
    \label{fig:all_ads_maps}
\end{figure*}

\end{document}